# A Comparative Analysis of Influenza Vaccination Programs

Shweta Bansal[1], Babak Pourbohloul[2], Lauren Ancel Meyers[3,4*]

1 Computational and Applied Mathematics, University of Texas Austin, Austin, Texas, United States of America, 2 UBC Centre for Disease Control, University of British Columbia, Vancouver, British Columbia, Canada, 3 Section of Integrative Biology and Institute for Cellular and Molecular Biology, University of Texas Austin, Austin, Texas, United States of America, 4 External Faculty, Santa Fe Institute, Santa Fe, New Mexico, United States of America

**Funding:** We acknowledge the financial support of the Canadian Institutes of Health Research, the Santa Fe Institute, and a NASA Harriett G. Jenkins Fellowship to SB. The funding agencies had no role in study design, data collection and analysis, decision to publish, or preparation of the manuscript.

**Competing Interests:** The authors have declared that no competing interests exist.

**Academic Editor:** Bryan Grenfell, Pennsylvania State University, United States of America

**Citation:** Bansal S, Pourbohloul B, Meyers LA (2006) A comparative analysis of influenza vaccination programs. PLoS Med 3(10): e387. DOI: 10.1371/journal.pmed.0030387

**Received:** October 13, 2005
**Accepted:** July 13, 2006
**Published:** October 3, 2006

**DOI:**
10.1371/journal.pmed.0030387

## ABSTRACT

### Background

The threat of avian influenza and the 2004–2005 influenza vaccine supply shortage in the United States have sparked a debate about optimal vaccination strategies to reduce the burden of morbidity and mortality caused by the influenza virus.

### Methods and Findings

We present a comparative analysis of two classes of suggested vaccination strategies: mortality-based strategies that target high-risk populations and morbidity-based strategies that target high-prevalence populations. Applying the methods of contact network epidemiology to a model of disease transmission in a large urban population, we assume that vaccine supplies are limited and then evaluate the efficacy of these strategies across a wide range of viral transmission rates and for two different age-specific mortality distributions.

We find that the optimal strategy depends critically on the viral transmission level (reproductive rate) of the virus: morbidity-based strategies outperform mortality-based strategies for moderately transmissible strains, while the reverse is true for highly transmissible strains. These results hold for a range of mortality rates reported for prior influenza epidemics and pandemics. Furthermore, we show that vaccination delays and multiple introductions of disease into the community have a more detrimental impact on morbidity-based strategies than mortality-based strategies.

### Conclusions

If public health officials have reasonable estimates of the viral transmission rate and the frequency of new introductions into the community prior to an outbreak, then these methods can guide the design of optimal vaccination priorities. When such information is unreliable or not available, as is often the case, this study recommends mortality-based vaccination priorities.

*The Editors' Summary of this article follows the references.*

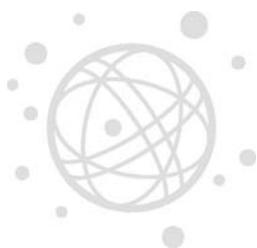

## Introduction

In response to the 2004–2005 influenza vaccine shortage, the United States Centers for Disease Control and Prevention (CDC) restricted vaccines to those most at risk for hospitalization and death — healthy infants, elderly individuals, and individuals with chronic illnesses. This strategy may be limited by the failure of vaccines to yield adequate protection for high-risk individuals [1,2] and the lesser roles played by infants and the elderly in disease transmission—they typically do not introduce influenza into households or other social groups.

Influenza outbreaks are believed to hinge, instead, on transmission by healthy school children [3–6], college students, and employed adults who have many daily contacts and are highly mobile [7]. Thus, epidemiologists have suggested an alternative approach: vaccinate school-age children to slow the spread of disease and thereby indirectly decrease mortality [8,9]. Several studies support this strategy. Monto et al. immunized school children in Tecumseh, Michigan, with inactivated influenza vaccine in 1968 and found lower total morbidity than in a matching community during a wave of influenza A (H3N2) [10]. Reichart et al. argue that mandatory influenza vaccination of school children in Japan from 1962 to 1987 reduced incidence and mortality among the elderly [11]. Recently, Longini et al. used mathematical models to show that, under certain assumptions, vaccinating 80% of all school-age children is almost as effective as vaccinating 80% of the entire population [8]. School-based vaccination programs have the additional benefits of high coverage, high efficacy, and minimal side effects [12].

In a similar spirit, others have suggested contact-based priorities that target individuals with the highest numbers of potentially disease-causing contacts [13–15]. This assumes that vulnerability is directly proportional to the number of contacts, and that removing the most vulnerable individuals from the transmission chain will maximally decrease disease spread. Identifying high-contact individuals in a community, however, may be difficult in practice.

Here we apply tools from contact network epidemiology [16–19] to evaluate vaccination strategies for a spectrum of influenza strains when vaccine supplies are limited. We use a realistic model of contact patterns in an urban setting to compare mortality-based strategies that target high-risk individuals to morbidity-based strategies that target demographics with high attack rates. We assess the efficacy of these measures for two substantially different virulence patterns, one based on mortality estimates from annual influenza epidemics and the other based on mortality estimates from the 1918 influenza pandemic. In addition, we consider the impact of vaccination delay and multiple imported cases on the relative effectiveness of the vaccination strategies.

## Methods

### Population Model

We built a contact network model that captures the interactions that underlie respiratory disease transmission within a city. The model is based on demographic information for Vancouver, British Columbia, Canada. In the model, each person is represented as a vertex, and interactions between people are represented as edges between appropriate vertices. Each person is assigned an age based on Vancouver census data, and age-appropriate activities (school, work, hospital, etc.). Interactions among individuals reflect household size, employment, school, and hospital data for Vancouver. The model population includes ~257,000 individuals. For further details and sensitivity analysis, see Protocol S1 and Figures S1 and S2.

Our contact network model contains undirected edges that reflect the possibility of disease transmission in either direction between two individuals, and directed edges that indicate the possibility of disease transmission from one person to another, but not the reverse (see Figure 1). Directed

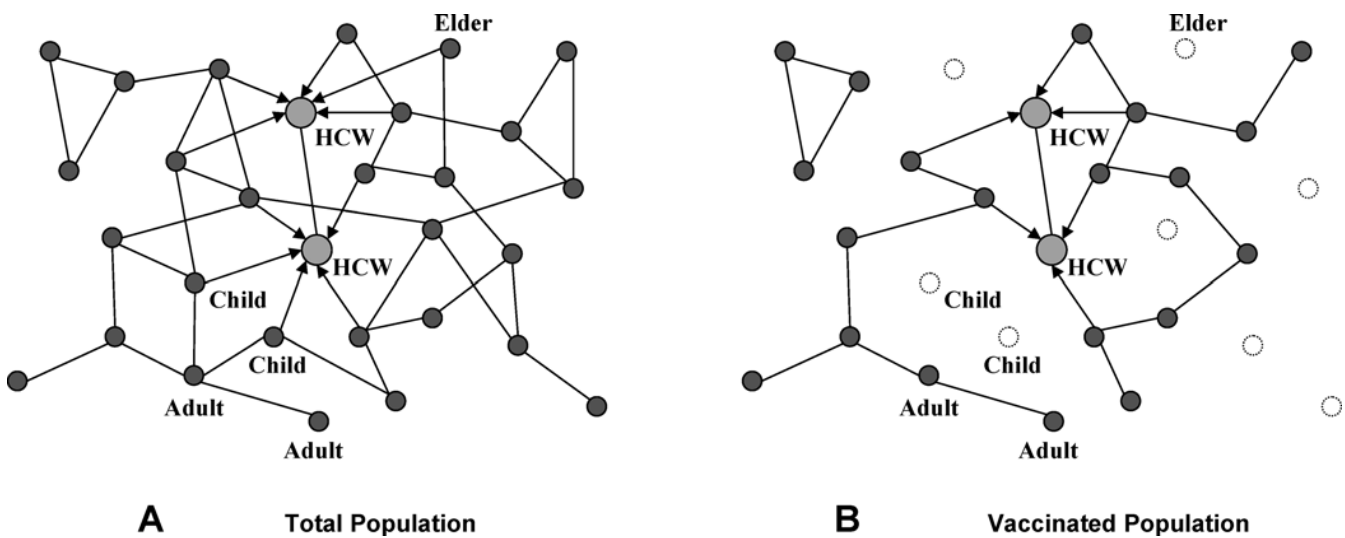

**Figure 1.** Network Model
(A) A schematic of a network model for an urban population. Each individual is a vertex in the network, and edges represent potentially disease-causing contacts between individuals. Directed edges (with arrows) represent transmission occurring in only one direction.
(B) We model vaccination in a population by removing nodes from the population network, and the edges that are attached to them.
HCW, health-care worker.
DOI: 10.1371/journal.pmed.0030387.g001

**Table 1.** The Age-Specific Mortality Distributions for Typical Annual Influenza Epidemics and an Example Influenza Pandemic

| Group | Mortality Rate for Influenza Epidemic (per 10,000 Cases) | Mortality Rate for Influenza Pandemic (per 10,000 Cases) |
| --- | --- | --- |
| Infants and toddlers (6 mo–3 y) | 0.30 | 80.0 |
| Preschool children (3–5 y) | 0.08 | 50.0 |
| Children (5–18 y) | 0.08 | 20.0 |
| Adults (18–50 y) | 0.07 | 70.0 |
| Elders (>50 y) | 12.00 | 5.0 |

DOI: 10.1371/journal.pmed.0030387.t001

edges model the possibility of transmission from an infected member of the general public to health-care workers during hospital visits. In a typical epidemic, most individuals infected with influenza do not seek hospital care. We assume that only high-risk groups (infants and elderly) visit hospitals upon infection and thus have opportunities to infect the health-care workers who treat them [20]. We also consider a more extreme scenario in which almost all infected individuals are at risk for serious complications and thus seek medical care upon infection.

### Influenza Mortality

Mortality rates differ both across demographic groups and among strains of influenza (see Table 1 and Protocol S1), and thus the optimal vaccination priorities are likely to depend on the virulence of the circulating strains. We consider two substantially different mortality models. The first assumes age-specific mortality rates typical of interpandemic outbreaks of flu, which are based on national viral surveillance data reported for 1977–1999 [21]. The rate is highest for the elderly, followed by infants, who are most at risk for death caused directly by influenza or pneumonia or by primary respiratory or circulatory complications. The second model, which was intentionally chosen for contrast, assumes mortality rates to be as estimated for the 1918 flu pandemic. These are high for healthy young adults aged 20–40 y and children under 5 y and low for older children and the elderly [22] (Table 1). There are, however, conflicting estimates for the elderly [23,24]. We use a low estimate to achieve the greatest departure from the interpandemic model, and thus to ascertain the sensitivity of our results to assumptions about influenza mortality. Henceforth, we refer to these two models as interpandemic and pandemic, respectively. We consider other reported mortality rates in Protocol S1 and Figure S3.

### Vaccine Priorities

We modeled targeted pre-season vaccination with single doses of inactivated influenza vaccine by removing select individuals (vertices) and all their contacts (edges) from the network before predicting the spread of influenza (see Figure 1). The fraction of the vaccinated population that becomes fully protected is based on demographic-specific vaccine efficacy estimates (Table 2). For a vaccine of efficacy $E$ and coverage $C$ for a particular group, we remove a fraction $C \cdot E$ of individuals from that group. This assumes that this fraction is fully protected (100% effectiveness) while the remaining fraction $C \cdot (1 - E)$ of vaccinated individuals are not protected at all. In reality, most of the vaccinated individuals will enjoy some partial protection. We have tested our method with simulations to confirm that it provides a reasonable model for partial efficacy (see Protocol S1 and Figure S5).

We evaluate four strategies (Figure 2): (1) a mortality-based strategy that, like the recent CDC strategy, targets demographics that are most vulnerable to health complications or death (infants, the elderly, and health-care workers for interpandemic flu; and infants, adults, and health-care workers for pandemic flu); (2) a morbidity-based strategy, similar to the priorities suggested by Longini et al [8] and Monto et al.[10], that targets school-aged children and school staff, and thereby aims to reduce mortality through herd protection [25]; (3) a mixed strategy that targets demographics with high attack rates (children) and high mortality

**Table 2.** Historical Influenza Vaccination Coverage Levels and Inactivated Vaccine Efficacy Levels Used in This Study

| Group | Vaccination Coverage Levels | Inactivated Vaccine Efficacy |
| --- | --- | --- |
| Infants and toddlers (6 mo–3 y) | 30%–75% | 70%–90% |
| Preschool children (3–5 y) | 30%–75% | 70%–90% |
| Children (5–18 y) | 30%–75% | 77%–91% |
| Adults (18–50 y) | 30%–75% | 70%–90% |
| Elders (>50 y) | 67%–85% | 30%–50% |
| Health-care workers | 25%–38% | 70%–90% |
| Elders in care facilities | 90%–95% | 30%–50% |

Sources: [12,26].
DOI: 10.1371/journal.pmed.0030387.t002

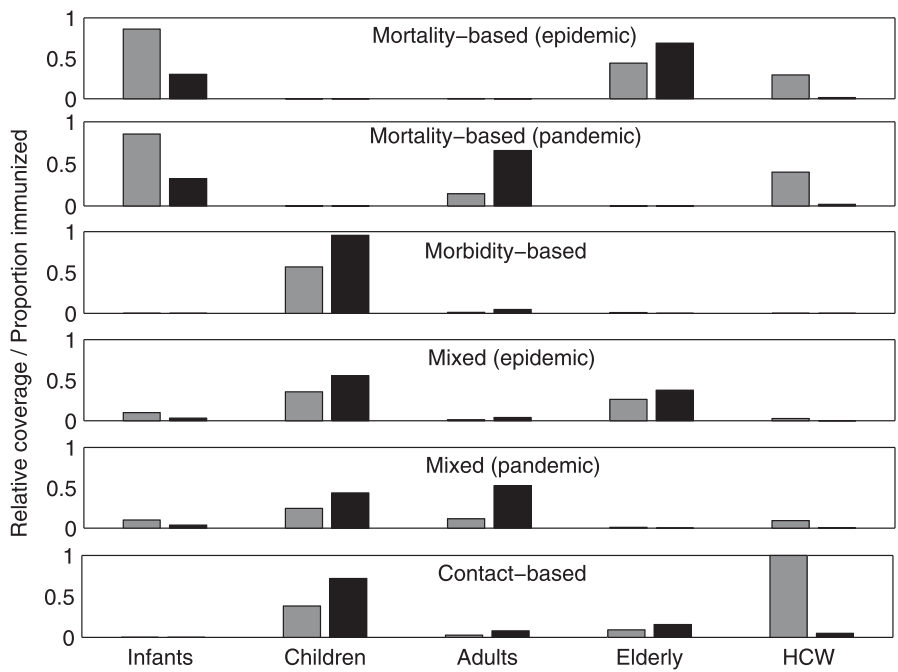

**Figure 2.** Vaccination Strategies
The demographic distribution of vaccines according to each of the strategies: the black bars reflect the fraction of available vaccines given to each age group (and thus will always sum to one). The gray bars reflect the proportion of each demographic that is effectively immunized, and thus take into account the size of the demographic and the demographic-specific vaccine efficacy.
HCW, health-care worker.
DOI: 10.1371/journal.pmed.0030387.g002

rates (infants and the elderly for interpandemic flu; infants and adults for pandemic flu), similar to that suggested by Longini and Halloran [9]; and (4) a contact-based strategy that removes a fraction of the most connected individuals.

We modeled the mortality-based strategy by removing infants, the elderly, and health-care workers from the network based on reported maximum coverage and efficacy levels for these demographics [12,26] (Table 2). This yielded 13% coverage of the total population (Table 3). We then implemented the remaining strategies to match this overall coverage level. (We consider the sensitivity of our results to the coverage level in Protocol S1 and Figure S4.) Targeted groups were removed in proportion to demographic-specific vaccine coverage levels reported in the 2002 National Health Interview Survey by the CDC [26], and the vaccine efficacy levels were based on age-specific rates reported for inactivated influenza vaccine [12].

### Epidemiological Analysis

We define the transmissibility of a disease, $T$, as the average probability that an infectious individual will transmit the disease to a susceptible individual with whom he or she has contact. This term summarizes important features of disease propagation including the contact rates among individuals, the duration of the infectious period, and the per contact probability of transmission. This per contact probability of transmission, in turn, summarizes the susceptibility (immune response) and the infectiousness (viral shedding) of individuals. Our analysis allows for variation in transmission rates from one individual to the next, but it assumes that these rates vary randomly with respect to the underlying contact patterns. There is evidence, however, that transmission rates may vary systematically among demographics, and, in particular, may be highest for children [27]. In Protocol S1 and Figure S6, we consider modified models that explicitly

**Table 3.** Vaccination Coverage and Efficacy Levels Assumed for the Mortality-Based Vaccination Strategy

| Group | Implemented Coverage Level | Vaccine Efficacy | Effective Coverage Level |
| --- | --- | --- | --- |
| Infants and toddlers (6 mo–3 y) | 75% (4.1% of total population) | 90% | 68% |
| Elders (>50 y) | 85% (7.5% of total population) | 50% | 43% |
| Health-care workers | 38% (0.4% of total population) | 90% | 34% |
| Elders in care facilities | 95% (0.7% of total population) | 50% | 48% |
| Total | 12.7% of total population | | |

The effective coverage level is a product of the implemented coverage level and the vaccine efficacy for each group.
DOI: 10.1371/journal.pmed.0030387.t003

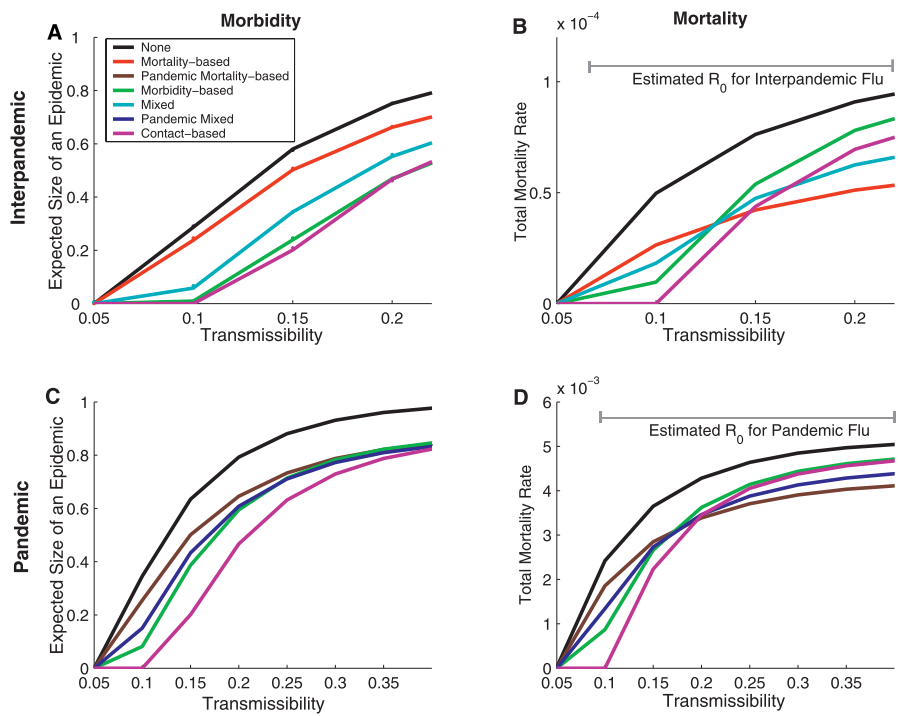

**Figure 3.** Morbidity and Mortality for Influenza Epidemics and Pandemics

Expected (A) attack rate and (B) mortality rate as a function of $T$ for annual influenza epidemics.

Expected (C) attack rate and (D) mortality rate as a function of $T$ for an influenza pandemic.

The dots in (A) show simulation results for comparison. Estimates of $R_0$ for interpandemic and pandemic flu are shown as gray lines in (B) and (D), respectively.

DOI: 10.1371/journal.pmed.0030387.g003

capture such demographic-specific variation in transmission rates and show that this additional complexity does not alter the results reported below.

$T$ is linearly related to the key epidemiological parameter $R_0$. In particular, $R_0$ is equal to $T \cdot \kappa$, where $\kappa$ is a measure of the connectivity within the population (network) [19,28]. Intuitively, $R_0$ is largest for highly contagious pathogens (represented by a high $T$) spreading through densely connected populations (represented by a high $\kappa$). $R_0 = 1$ corresponds to a critical transmissibility value $T_c$, above which a population is vulnerable to large-scale epidemics and below which only small outbreaks occur [28].

We used methods based on contact network epidemiology [16–19] to predict the fate of an influenza outbreak as a function of the average transmissibility $T$ of the strain. For any contact network, one can mathematically predict the epidemic threshold ($T_c$), the average size of a small outbreak ($s$), the average size ($S_e$) and probability ($P_e$) of a large-scale epidemic, and demographic-specific attack rates for an epidemic, should one occur. Mortality is predicted by multiplying the expected number of infections for a given group by the age-specific mortality rate assumed for that group. (See Protocol S1 for additional details.)

To verify these mathematical predictions, we performed numerical simulations of disease spread assuming a simple susceptible–infectious–recovered (SIR) model. Beginning with a susceptible network and a single infected case, we iteratively take each currently infected vertex, infect each of its susceptible contacts with probability $T$, and then change the status of the original vertex to "recovered." These simulations are generally consistent with the mathematical calculations (Figure 3), and thus we primarily report the analytical results.

Immunity from prior outbreaks is an important aspect of interpandemic influenza transmission. There are two alternative approaches to modeling residual immunity. One is to remove individuals with naturally acquired immunity from the network, as we have done for vaccination. The other is to assume that the distribution of transmission probabilities reflects pre-existing immunity. If there is widespread partial immunity, then there will be large numbers of edges along which transmission is very unlikely, leading to a lower average transmissibility across the population. Here we have not removed individuals with naturally acquired immunity from the population, but instead assume that the transmissibility values are averaged over all edges in the network, including those leading to or from such individuals.

### Model Validation

We compared the age-specific attack rates predicted by our models to those reported for both interpandemic flu and the 1918 pandemic (Figure 4). First, we considered data from the interpandemic outbreak of 1977–1978 reported by Longini et al. [3]. They reported age-specific attack rates from a household study of 159 families in Seattle, Washington, United States, in which infection was determined through hemagglutination-inhibition assays. We do not know the exact age-specific influenza vaccination coverage rates during this period. We assumed that the population was protected according the current CDC strategy (the mortality-based strategy), and then used our model to predict demographic-specific attack rates. We estimated the average transmissi-bility of the disease by solving for the value of $T$ that produces the observed total attack rate ($T = 0.07$, or $R_0 = 1.2$). Thus, the total attack rate was constrained to match perfectly the observed total attack rate, while the demographic-specific attack rates were free to vary. The predictions of the model are consistent with the observed epidemiology (Figure 4A). We note, however, that the reliability of this comparison is limited by the small sample size of the Seattle study and the lack of information about vaccine coverage and efficacy during that period.

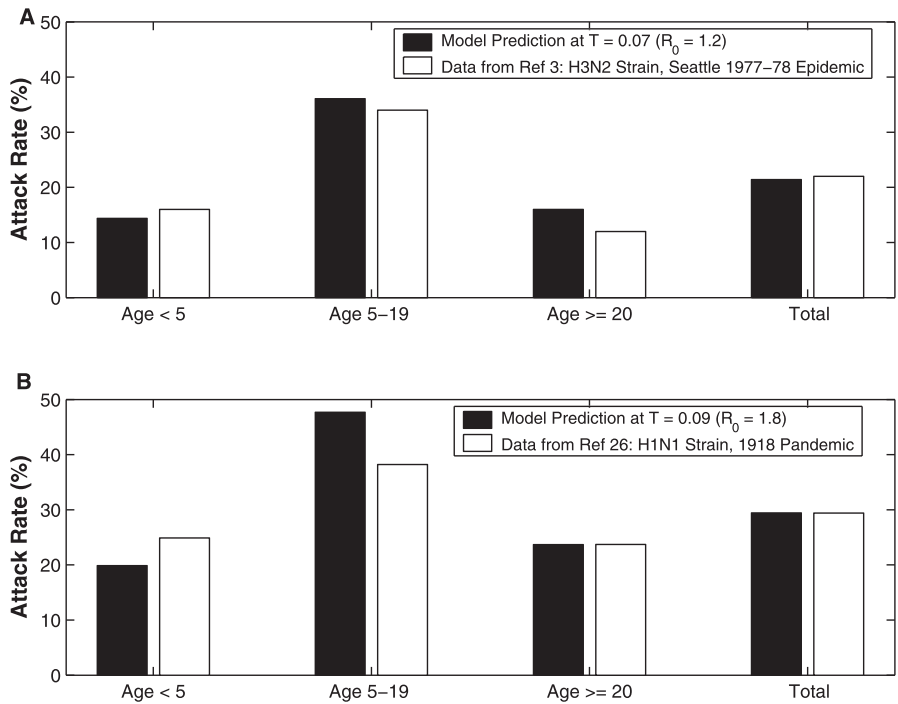

**Figure 4.** Model Validation
Comparison of predicted to observed age-specific attack rates for (A) the 1977–1978 influenza season and (B) the 1918 influenza pandemic.
DOI: 10.1371/journal.pmed.0030387.g004

Second, we made a similar comparison using age-specific attack rate data for the 1918 pandemic that were collected and reported by Frost in 1920 [29]. The data are based on a survey of approximately 146,000 people (representing a cross-section of the US population, which at the time numbered 103 million). Infection rates for influenza were based on self-reported responses by study participants. There was no vaccination available for influenza at the time in the US, and thus we made epidemiological predictions assuming no vaccination. Again, we began by solving for an average transmissibility that produces the observed total attack rate and found $T = 0.09$ (or $R_0 = 1.8$). As a consistency check, this estimate agrees very closely with the recently revised estimate for the pandemic influenza reproductive rate [30], based on US and UK 1918 pandemic mortality data. Assuming this average transmissibility, we predicted demographic-specific attack rates and found that they matched the observed patterns reasonably well (Figure 4B).

## Results/Discussion

### Direct versus Indirect Intervention Methods

For interpandemic influenza, morbidity-based and contact-based strategies appear to offer significant indirect protection of unvaccinated individuals who would otherwise become infected via transmission chains that have now been severed by vaccination. Indeed, for all strains, these two strategies are predicted to yield the lowest attack rates (Figure 3A). If the primary objective is to reduce morbidity from influenza, then the morbidity-based and contact-based strategies are always preferred, although their advantage decreases as disease transmissibility $(T)$ increases.

One might argue that the primary objective of intervention should be to reduce mortality rather than morbidity. The CDC's recent vaccine priorities seem to be based on this objective [12]. In terms of mortality, there is a specific transmissibility value below which the morbidity-based and contact-based strategies are superior and above which the mortality-based strategies are superior (Figure 3B). To clarify this transition (which occurs for our network at $T = 0.13$), we show in Figure 5 the proportions of the adult and elderly subpopulations that are infected, vaccinated, and uninfected for the two strategies at two values of $T$. The uninfected class is made up of individuals that have neither been vaccinated nor get infected. Some of these individuals would not be infected in any case, and the rest are those that would be infected without a vaccination program but are now protected by the effects of herd immunity. Below the transition point (for instance, at $T = 0.1$), the elderly are protected more by the indirect effects of the morbidity-based strategy than by the direct effects of the mortality-based strategy. Above the transition point (for instance, at $T = 0.15$), the indirect protection by the morbidity-based strategy drops substantially, resulting in a higher proportion of elderly individuals infected than with the mortality-based strategy. A similar reversal occurs for infants. The mixed strategy—a combination of the morbidity-based and mortality-based strategies—is never the optimal strategy (Figure 3B), yet may be an advisable bet-hedging strategy when there is great uncertainty about the transmissibility of the circulating strain.

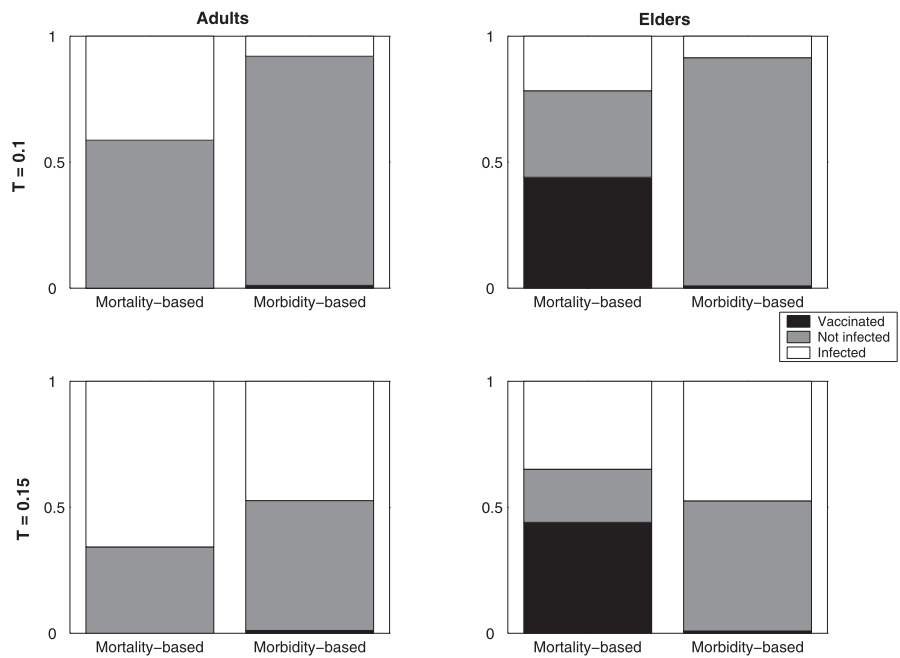

**Figure 5.** Direct versus Indirect Intervention
The figure shows the proportions of the adult and elderly populations that are infected, not infected (neither vaccinated nor infected), and vaccinated for two different values of $T$ for the mortality-based strategy versus the morbidity-based strategy.
DOI: 10.1371/journal.pmed.0030387.g005

Estimates of $R_0$ for interpandemic flu range between 1.0 and 2.4 for the A (H2N2) and A (H3N2) strains of influenza ([31,32] and references therein). Since influenza vaccines have been used in the US since 1944, these estimates may be based on partially vaccinated populations. Conservatively assuming that the populations in question had somewhere between no coverage at all and 13% coverage according to the contact-based strategy, these values of $R_0$ ($1.0 < R_0 < 2.4$) correspond to $0.06 < T < 0.26$ in our model (See Protocol S1.) This range straddles the critical cross-points in Figure 3B, leaving some ambiguity as to which strategy will be most effective. We note, however, that the higher the transmissibility, the more dire the public health situation, and mortality-based strategies are predicted to be more effective for highly contagious strains.

### Highly Virulent Influenza

The demographic-specific mortality rates reported for influenza vary considerably (Protocol S1 and Figure S3). To assess whether control recommendations can be generalized to new or anomalous strains of influenza, we analyzed a second, extreme scenario. Worldwide influenza pandemics are characterized by much higher levels of morbidity and mortality than annual epidemics, and have occurred three times in the last century. The 1918–1919 "Spanish Influenza" caused more than 500,000 deaths in the US and an estimated 20 million deaths worldwide [33]. Based on data from the 1918 pandemic, we modified our model in three respects: the number of people expected to seek medical attention upon infection, the age-specific mortality rates, and (consequentially) the age groups targeted by the mortality-based and mixed strategies.

Despite these substantial differences, the predictions for pandemic and interpandemic flu are qualitatively similar. The morbidity-based and contact-based strategies outperform mortality-based strategies in terms of resulting mortality for low values of $T$, but not for higher values. There is a quantitative difference, however, in that the transition point between these two regimes happens at a higher transmissibility for pandemic flu than for interpandemic flu (Figure 3D versus 3B). In other words, morbidity-based strategies are preferred for a wider spectrum of pandemic flu strains than of interpandemic flu strains. This stems, in part, from the much larger size of the high-risk population (adults) for pandemic flu. Under vaccine limitations (13% in this case), the mortality-based strategy protects a much smaller fraction of the pandemic high-risk population than of the interpandemic high-risk population. We have found that increasing the vaccination level to 20% does not change the qualitative results (shown in Protocol S1 and Figure S4). Patel et al. have recently performed a similar sensitivity analysis on vaccine availability [34].

The reproductive number ($R_0$) for the 1918 Spanish Influenza is estimated to have been between 1.8 and 4.0 [29,35], corresponding to $T$ between 0.09 and 0.43 in our model (see Protocol S1). Once again, this range straddles the critical cross-point in Figure 3D, leaving some ambiguity as to which strategy will be most effective. It can be seen, however, that mortality-based strategies are predicted to be more effective across the upper two-thirds of this interval.

### Multiple Introductions

Most communities do not exist in isolation, and thus experience multiple independent introductions of the virus during a typical flu season. Many models of vaccination strategies [8,9], however, ignore this possibility. To better understand the probability and rates of new importations of flu, one must consider a meta-population model that includes connectivity among cities. Here we address the consequences of multiple introductions, but not the likelihood of such events in the first place. For mathematical simplicity, we assumed that multiple independent introductions occur simultaneously (and initial cases are chosen randomly) at the start of an outbreak, which yields conservative estimates of their detrimental impact. The probability of an epidemic increases with the number of introductions for all strategies, thereby reducing the advantage of the morbidity-based and contact-based strategies for mildly transmissible strains. For example, if there are four independent introductions of flu, morbidity-based strategies are inferior to mortality-based strategies above $T = 0.12$ ($R_0 = 2.1$). In contrast, this shift takes place at $T = 0.13$ ($R_0 = 2.3$), when there is a single importation of disease (Figure 6).

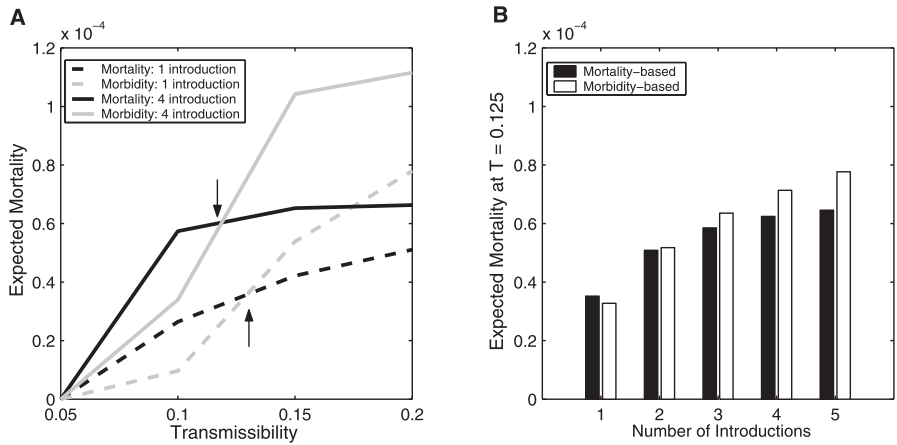

**Figure 6.** The Epidemiological Impact of Multiple Introductions of Disease
(A) The morbidity-based strategy is more effective than the mortality-based strategy when $T$ is less than 0.13 only if there is only a single introduction of disease. With four introductions of disease, however, the morbidity-based strategy becomes less effective (and is the preferred strategy only when $T$ is less than 0.12.)
(B) At $T = 0.125$, the morbidity-based strategy is superior to the mortality-based strategy when there is a single introduction, but inferior when there is more than one introduction.
DOI: 10.1371/journal.pmed.0030387.g006

### Delayed Intervention

A similar analysis provides insight into the impact of a delay in intervention until after an outbreak is already in progress, as occurred during the 2000–2001 flu season [36]. This scenario may also be particularly relevant to pandemic influenza, for which vaccines may only become available well into an outbreak, if at all. We simulate the implementation of vaccination after a certain proportion of the population has already been infected. We call this proportion "delay." The morbidity-based strategies are more sensitive to such delays than mortality-based methods are (Figure 7). They are predicted to be inferior above $T = 0.11$ ($R_0 = 1.9$) if there is a 10% delay in vaccination, compared to $T = 0.13$ ($R_0 = 2.3$) when there is no delay.

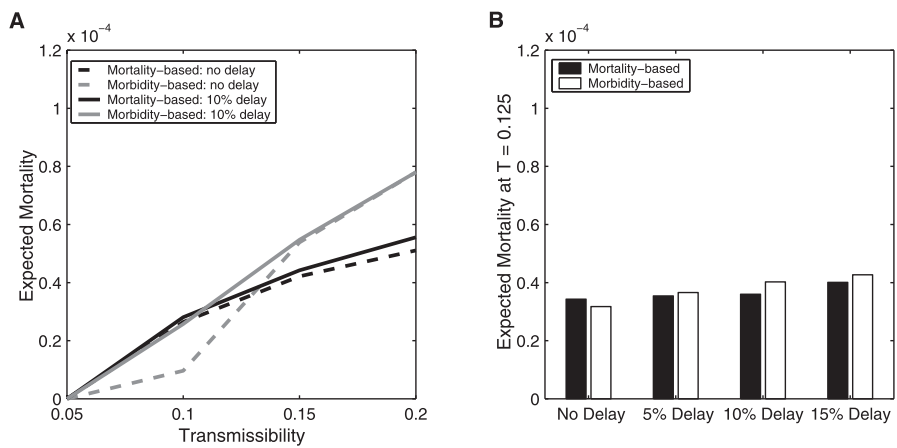

**Figure 7.** The Epidemiological Impact of Delayed Vaccination
(A) The morbidity-based strategy is more effective than the mortality-based strategy when $T$ is less than 0.13 only if there is there is no delay in vaccination. When vaccines are given after 10% of the population has already been infected, the morbidity-based strategy becomes relatively less effective (and is the preferred strategy only when $T$ is less than 0.11).
(B) At $T = 0.125$, the morbidity-based strategy is superior to the mortality-based strategy when there is no delay, but inferior for any amount of delay. Each of the values is an average taken across 500 epidemic simulations on the contact network.
DOI: 10.1371/journal.pmed.0030387.g007

Figures 6 and 7 suggest that a delay in vaccination may be less detrimental than multiple introductions of disease into a population. Multiple independent introductions of disease provide multiple independent opportunities to spark a large-scale epidemic. In the absence of vaccination, the probability of an epidemic increases considerably as the number of independent introductions increases (Protocol S1). In contrast, a delay in vaccination allows a single case to grow into a connected cluster of cases, which are not independent of each other with respect to the numbers and the identities of their contacts. The probability of an epidemic increases with the number of individuals in the initial cluster, but not as quickly as it does with the addition of independent cases.

### Conclusion

In this study, we have applied the analytical methods of contact network epidemiology to evaluate current and proposed influenza vaccination priorities. In contrast to prior studies [9,34], we have modeled a relatively large population and the entire spectrum of viral transmission rates possible for influenza; in addition, we have accounted for multiple introductions of disease and the possibility of a delay in vaccination. The efficacy of mortality-based strategies (like the CDC 2004 vaccination priorities [12]) and morbidity-based strategies (like school-based vaccination [8,9]) depend on (i) the transmissibility (reproductive number) of the strain, (ii) age-specific mortality rates, (iii) the vulnerability of the community to multiple introductions, and (iv) the timing of implementation. With respect to minimizing mortality, mortality-based strategies are generally preferred to morbidity-based strategies for strains with high transmission rates and in communities experiencing either delayed intervention or multiple introductions.

Thus, mortality-based strategies may be the prudent choice for outbreaks of new or atypical strains of influenza, when public health officials may not have reliable estimates for all (or any) of the first three inputs, and vaccination may be delayed. The predictions appear to hold for a range of age-specific mortality distributions estimated for past outbreaks of epidemic and pandemic flu. Although this suggests that similar recommendations may be appropriate for pandemic flu, they will be irrelevant in the very likely case that vaccines are not available at the start of an outbreak.

If more precise estimates of the key inputs become available, then this approach can be applied to design optimal (rather than just prudent) priorities. To reduce the existing uncertainty in estimates of influenza transmission and mortality rates, we must improve surveillance methods for gathering real-time data and develop new statistical methods for examining data from both historical and future outbreaks, as were developed to estimate $R_0$ for SARS during the 2003 outbreaks [37]. Current estimates of flu transmission rates are based primarily on compartmental models of disease transmission ([31,32] and references therein). Some of the variation in the estimates of $R_0$ may stem from variation in contact patterns among different populations rather than intrinsic variation in the probability of disease transmission between individuals who come in contact with one another. While compartmental models often do not capture such contact heterogeneity, contact network models allow one to factor out variation in contact patterns when estimating transmission rates. Thus, the development of better estimation methods using contact network models may yield more accurate estimates of some key epidemiological parameters.

### Supporting Information

**Figure S1.** (Normalized) Degree Distributions for Various Demographic Groups before and after Vaccination (with the Interpandemic Mortality-Based Strategy and the Morbidity-Based Strategy)

Found at DOI: 10.1371/journal.pmed.0030387.sg001 (65 KB PDF).

**Figure S2.** Variation in the Size of Epidemic and Total Mortality Predicted for Mortality-Based and Morbidity-Based Strategies across 100 Networks with 100% Variation in Contact Parameters

Found at DOI: 10.1371/journal.pmed.0030387.sg002 (64 KB PDF).

**Figure S3.** Epidemiological Predictions for Five Different Estimated Influenza Mortality Distributions

Found at DOI: 10.1371/journal.pmed.0030387.sg003 (67 KB PDF).

**Figure S4.** Total Mortality at a 20% Vaccination Coverage Level

Found at DOI: 10.1371/journal.pmed.0030387.sg004 (61 KB PDF).

**Figure S5.** Results from Simulation Demonstrate that the Two Methods of Modeling Vaccine Efficacy Give Similar Results

Found at DOI: 10.1371/journal.pmed.0030387.sg005 (65 KB PDF).

**Figure S6.** Results for Total Mortality Rate with Variation in Infectivity and Susceptibility

Found at DOI: 10.1371/journal.pmed.0030387.sg006 (58 KB PDF).

**Protocol S1.** Quantitative Comparison of Influenza Vaccination Programs

Found at DOI: 10.1371/journal.pmed.0030387.sd001 (159 KB PDF).

### Acknowledgments

The authors would like to thank Martin Meltzer, Robert C. Brunham, Mel Krajden, Danuta M. Skowronski, and James Lloyd-Smith for their insightful suggestions. We also thank three anonymous referees for their valuable comments and suggestions.

**Author contributions.** All three authors contributed to the design, analysis, and presentation of this work.

### References

1. Govaert TME, Thijs CT, Masurel N, Sprenger MJ, Dinant GJ, et al. (1994) The efficacy of influenza vaccination in elderly individuals. JAMA 272: 1661–1665.
2. Brandriss MW, Betts RF, Mathur U, Douglas RG Jr (1981) Responses of elderly subjects to monovalent A(HINI) and trivalent A(HINI)/A(H3N2)/B vaccines. Am Rev Respir Dis 124: 681–684.
3. Longini IM, Koopman JS, Monto AS, Fox JP (1982) Estimating household and community transmission parameters of influenza. Am J Epidemiol 115: 736–751.
4. Fox JP, Hall CE, Cooney MK, Foy HM (1982) Influenza virus infections in Seattle families, 1975–1979. Am J Epidemiol 116: 212–227.
5. Jennings LC, Miles JAR (1978) A study of acute respiratory disease in the community of Port Chalmers. J Hyg (Lond) 81: 67–75.
6. Taber LH, Paredes A, Glezen WP, Couch RB (1982) Infection with influenza A/Victoria virus in Houston families. J Hyg (Lond) 86: 303–313.
7. Glezen WP (1996) Emerging infections: Pandemic influenza. Epidemiol Rev 18: 64–76.
8. Longini IM Jr, Halloran ME, Nizam A, Yang Y (2004) Containing pandemic influenza with antiviral agents. Am J Epidemiol 159: 623–633.
9. Longini IM, Halloran ME (2005) Strategy for distribution of influenza vaccine to high-risk groups and children. Am J Epidemiol 161: 303–306.
10. Monto AS, Koopman JS, Longini IM Jr (1985) The Tecumseh study of illness. XIII. Influenza infection and disease. Am J Epidemiol 121: 811–822.
11. Reichart TA, Sugaya N, Fedson DS, Glezen WP, Simonsen L, et al. (2001) The Japanese experience with vaccinating school-children against influenza. N Engl J Med 344: 889–896.
12. Harper SA, Fukuda K, Uyeki TM, Cox NJ, Bridges CB, et al. (2004) Prevention and control of influenza: Recommendations of the Advisory Committee on Immunization Practices (ACIP). MMWR 53: 1–40.
13. Dezso Z, Barabasi AL (2001) Halting viruses in scale-free networks. Phys Rev E Stat Nonlin Soft Matter Phys 65: 055103.

14. Pastor-Satorras R, Vespignani A (2001) Immunization of complex networks. Phys Rev E Stat Nonlin Soft Matter Phys 65: 036104.
15. Woolhouse ME, Dye C, Etard JF, Smith T, Charlwood JD, et al. (1997) Heterogeneities in the transmission of infectious agents: Implications for the design of control programs. Proc Natl Acad Sci U S A 94: 338–342.
16. Meyers LA, Newman MEJ, Martin M, Schrag S (2003) Applying network theory to epidemics: Control measures for *Mycoplasma pneumoniae* outbreaks. Emerg Infect Dis 9: 204–210.
17. Meyers LA, Pourbohloul B, Newman MEJ, Skowronski DM, Brunham RC (2005) Network theory and SARS: Predicting outbreak diversity. JTB. 232: 71–81.
18. Pourbohloul B, Meyers LA, Skowronski DM, Krajden M, Patrick DM, et al. (2005) Modeling Control Strategies of Respiratory Pathogens. Emerg Infect Dis. 11: 1249–1256
19. Meyers LA, Newman MEJ, Pourbohloul B (2006) Predicting epidemics on directed contact networks. J Theor Biol 240: 400–418.
20. Thompson WW, Shay DK, Weintraub E (2004) Influenza-associated hospitalizations in the United States. JAMA 292: 1333–1340.
21. Thompson WW, Shay DK, Weintraub E, Brammer L, Cox N, et al. (2003) Mortality associated with influenza and respiratory syncytial virus in the United States. JAMA 289: 179–186.
22. Luk J, Gross P, Thompson WW (2001) Observations on mortality during the 1918 influenza pandemic. Clin Infect Dis 33: 1375–1378.
23. Dauer CC, Serfling RE (1961) Mortality from influenza, 1957–1958 and 1959–1960. Am Rev Respir Dis 83: 15–26.
24. Olsen DR, Simonsen L, Edelson PJ, Morse SS (2005) Epidemiological evidence of an early wave of the 1918 influenza pandemic in New York City. Proc Natl Acad Sci U S A 102: 11059–11063.
25. Paul Y (2004) Herd immunity and herd protection. Vaccine 22: 301–302.
26. Centers for Disease Control and Prevention (2005) Interim estimates of populations targeted for influenza vaccination from 2002 National Health Interview Survey data and estimates for 2004 based on influenza vaccine shortage priority groups. Atlanta (Georgia): Centers for Disease Control and Prevention. 2 p.
27. Cauchemez S, Carrat F, Viboud C, Valleron AJ, Boelle PY (2004) A Bayesian MCMC approach to study transmission of influenza: Application to household longitudinal data. Stat Med 23: 3469–3487.
28. Newman MEJ (2002) Spread of epidemic disease on network. Phys Rev E Stat Nonlin Soft Matter Phys 66: 016128.
29. Frost WH (1920) Statistics of influenza morbidity: With special reference to certain factors in case incidence. Public Health Rep 36: 584–597.
30. Ferguson NM, Cummings DA, Cauchemez S, Fraser C, Riley S, et al. (2005) Strategies for containing an emerging influenza pandemic in Southeast Asia. Nature 437: 209–214.
31. Hethcote HW (2000) The mathematics of infectious diseases. SIAM Rev 42: 599–653.
32. Grais RF, Ellis JH, Kress A, Glass GE (2004) Modeling the spread of annual influenza epidemics in the U.S.: The potential role of air travel. Health Care Manag Sci 7: 127–134.
33. Noble GR (1982) Epidemiological and clinical aspects of influenza. In: Beare AS, editor. Basic and applied influenza research. Boca Raton (Florida): CRC Press. pp. 11–50.
34. Patel R, Longini IM, Halloran ME (2005) Finding optimal vaccination strategies for pandemic influenza using genetic algorithms. J Theor Biol 234: 201–212.
35. Mills CE, Robins JM, Lipsitch M (2004) Transmissibility of 1918 pandemic influenza. Nature 432: 904–906.
36. Centers for Disease Control and Prevention (2000) Flu season 2000–01: Flu vaccine supply. Atlanta (Georgia): Centers for Disease Control and Prevention. Available: http://www.cdc.gov/od/oc/media/pressrel/r2k0622a.htm. Accessed 25 March 2005.
37. Cauchemez S, Boelle PY, Donnelly CA, Ferguson NM, Thomas G, et al. (2006) Real-time estimates in early detection of SARS. Emerg Infect Dis 12: 110–113.

### Editors' Summary

**Background.** Influenza—a viral infection of the nose, throat, and airways that is transmitted in airborne droplets released by coughing or sneezing—is a serious public health threat. Most people recover quickly from influenza, but some individuals, especially infants, old people, and individuals with chronic health problems, can develop pneumonia and die. In the US, seasonal outbreaks (epidemics) of flu cause an estimated 36,000 excess deaths annually. And now there are fears that avian influenza might start a human pandemic—a global epidemic that could kill millions. Seasonal outbreaks of influenza occur because flu viruses continually change the viral proteins (antigens) to which the immune system responds. "Antigenic drift"—small changes in these proteins—means that an immune system response that combats flu one year may not provide complete protection the next winter. "Antigenic shift"—large antigen changes—can cause pandemics because communities have no immunity to the changed virus. Annual vaccination with vaccines based on the currently circulating viruses controls seasonal flu epidemics; to control a pandemic, vaccines based on the antigenically altered virus would have to be quickly developed.

**Why Was This Study Done?** Most countries target vaccination efforts towards the people most at risk of dying from influenza, and to health-care workers who are likely come into contact with flu patients. But is this the best way to reduce the burden of illness (morbidity) and death (mortality) caused by influenza, particularly at the start of a pandemic, when vaccine would be limited? Old people and infants are much less likely to catch and spread influenza than school children, students, and employed adults, so could vaccination of these sections of the population—instead of those most at risk of death—be the best way to contain influenza outbreaks? In this study, the researchers used an analytical method called "contact network epidemiology" to compare two types of vaccination strategies: the currently favored mortality-based strategy, which targets high-risk individuals, and a morbidity-based strategy, which targets those segments of the community in which most influenza cases occur.

**What Did the Researchers Do and Find?** Most models of disease transmission assume that each member of a community is equally likely to infect every other member. But a baby is unlikely to transmit flu to, for example, an unrelated, housebound elderly person. Contact network epidemiology takes the likely relationships between people into account when modeling disease transmission. Using information from Vancouver, British Columbia, Canada, on household size, age distribution, and occupations, and other factors such as school sizes, the researchers built a model population of a quarter of a million interconnected people. They then investigated how different vaccination strategies controlled the spread of influenza in this population. The optimal strategy depended on the level of viral transmissibility—the likelihood that an infectious person transmits influenza to a susceptible individual with whom he or she has contact. For moderately transmissible flu viruses, a morbidity-based vaccination strategy, in which the people most likely to catch the flu are vaccinated, was more effective at containing seasonal and pandemic outbreaks than a mortality-based strategy, in which the people most likely to die if they caught the flu are vaccinated. For highly transmissible strains, this situation was reversed. The level of transmissibility at which this reversal occurred depended on several factors, including whether vaccination was delayed and how many times influenza was introduced into the community.

**What Do These Findings Mean?** The researchers tested their models by checking that they could replicate real influenza epidemics and pandemics, but, as with all mathematical models, they included many assumptions about influenza in their calculations, which may affect their results. Also, because the contact network used data from Vancouver, their results might not be applicable to other cities, or to nonurban areas. Nevertheless, their findings have important public health implications. When there are reasonable estimates of the viral transmission rate, and it is known how often influenza is being introduced into a community, contact network models could help public health officials choose between morbidity- and mortality-based vaccination strategies. When the viral transmission rate is unreliable or unavailable (for example, at the start of a pandemic), the best policy would be the currently preferred strategy of mortality-based vaccination. More generally, the use of contact network models should improve estimates of how infectious diseases spread through populations and indicate the best ways to control human epidemics and pandemics.

**Additional Information.** Please access these Web sites via the online version of this summary at http://dx.doi.org/10.1371/journal.pmed.0030387.

- US Centers for Disease Control and Prevention information about influenza for patients and professionals, including key facts on vaccination
- US National Institute of Allergy and Infectious Diseases feature on seasonal, avian, and pandemic influenza
- World Health Organization fact sheet on influenza, with links to information on vaccination
- UK Health Protection Agency information on seasonal, avian, and pandemic influenza
- MedlinePlus entry on influenza

# Supporting Information:

Quantitative Comparison of Influenza Vaccination Programs

Shweta Bansal, Babak Pourbohloul, Lauren Ancel Meyers

SUPPLEMENTAL METHODS
*Urban Contact Network Generation*

For this study, we generate plausible contact networks for an urban setting using demographic information for the Greater Vancouver Regional District, which is the third largest metropolitan area in Canada. We use publicly available data from sources such as Statistics Canada to estimate the distribution of ages, household sizes, school and classroom sizes, hospital occupancy, workplaces, and public spaces [S1-S4]. Qualitatively similar age and household size distributions are found for other cities in Canada ranging in population sizes from 120,000 to 4.6 million [S5]. We begin assembling the urban network by choosing 100,000 households at random from the Vancouver household size distribution [S1], which yields approximately 257,000 people according to a mean household size of approximately 2.6. Based on ages assigned from the measured Vancouver age distribution [S2], each member of the population is assigned to activities: to schools according to school and class size distributions [S3]; to occupations according to (un)employment data; to hospitals as patients and caregivers according to hospital employment and bed data [S4]; to nursing homes according to nursing home occupancy data; and to other public places.

To model heterogeneities in contact patterns, we create random connections (edges) between individuals (nodes) based on the location and nature of their overlapping daily activities. Individuals in households are connected with probability 1, while individuals encountering others in public places are connected with probabilities ranging from 0.003 to 0.3. Each school and hospital is subdivided into classrooms or wards. Pairs of students and pairs of patients within these subunits are connected with higher probability than pairs associated with different subunits. Teachers are assigned to classrooms and connected stochastically to appropriate students. Caregivers are assigned wards and then connected to appropriate patients. There are also low probability neighborhood contacts between individuals from different households.

*Epidemiological Analysis*

The methods described in this section are derived and described fully in Ref S6. Here, we only present the epidemiological equations with a few motivating details. Our contact network models are *semi-directed* – containing both undirected and directed edges. In a semi-directed network, each vertex (individual) has an *undirected-degree* representing the number of undirected edges joining the vertex to other vertices as well as both an *in-degree* and an *out-degree* representing the number of directed edges coming *from* other vertices and going *to* other vertices, respectively. The undirected-degree and in-degree indicate how many contacts can spread disease to the individual, and thus is related to the likelihood that an individual will become infected during an epidemic; and the undirected-degree and out-degree indicate how many contacts may be infected by that individual should he or she become infected, and thus is related to the likelihood that an individual will ignite an epidemic.

Given the degree distribution of the contact network within a population, one can analytically predict what will happen when an infectious disease like influenza enters the population. Let $p_{jkm}$ be the probability that any given person in the population has in-degree equal to $j$, out-degree equal to $k$, and undirected-degree $m$. Let $T$ be the transmissibility of the disease, that is, the average probability that transmission of the disease occurs between an infected individual and a susceptible individual with whom they are in contact.

Network theory makes a technical distinction between outbreaks and epidemics. An outbreak is a causally connected cluster of cases which, by chance or because the transmission probability is low, dies out before spreading to the population at large. In an epidemic, on the other hand, the infection escapes the initial group of cases into the community at large and results in population-wide incidence of the disease. The crucial difference is that the size of an outbreak is determined by the spontaneous dying out of the infection, whereas the size of an epidemic is limited only by the size of the population through which it spreads.

To predict the fate of an outbreak, we use *probability generating functions*, to summarize useful information about network topology. Thus, if a graph has degree distribution $p_{jkm}$, then the probability generating function (PGF, henceforth) for $p_{jkm}$ is

$$\Gamma(x, y, u) = \sum_{jkm} p_{jkm} x^j y^k u^m.$$

The average in-degree, out-degree, and undirected-degree are equal to:
$\langle k_{in} \rangle = \sum_{jkm} j p_{jkm}$, $\langle k_{out} \rangle = \sum_{jkm} k p_{jkm}$, and $\langle k_{un} \rangle = \sum_{jkm} m p_{jkm}$. The average degrees can also be attained by evaluating the partial derivatives of $\Gamma(x, y, u)$ at $x = 1, y = 1$ and $u = 1$. We note that since every directed edge has an origin and a destination, the average in-degree equals the average out-degree $\left( \sum_{jkm} j p_{jkm} = \sum_{jkm} k p_{jkm} = \sum_{jkm} \frac{j+k}{2} p_{jkm} \right)$.

If you choose a random *directed* edge in the network and follow it to the nearest vertex, then the PGF for the number of the three types of edges (in, out, and undirected) emanating from that vertex other than the one that we arrived on is

$$H_d(x,y,u) = \frac{\sum_{jkm} j p_{jkm} x^{j-1} y^k u^m}{<k_{in}>}.$$

Likewise, if you choose a random *undirected* edge in the network and follow it to the nearest vertex, then the PGF for the various edges at that vertex is given by

$$H_u(x,y,u) = \frac{\sum_{jkm} m p_{jkm} x^j y^k u^{m-1}}{<k_{un}>}.$$

Using these methods, we can derive the reproductive ratio, $R_0$, the average size of an outbreak, $\langle s \rangle$, the size of an epidemic, $S_e$, the probability of an epidemic, $P_e$, and the probability that an individual with a certain (in- and undirected-) degree will become infected, $v_{jm}$.

*The basic reproductive ratio*: When calculating the expected number of new cases arising from an infection in a naïve population we consider the source vertex of the infection. That is, the initial case may arise through infection along a directed or undirected edge. Thus, if we know the source of the infection we can more accurately predict the $R_0$. In particular,

$$R_0^d = T \frac{\sum_{jkm} j(k+m) p_{jkm}}{<k_{in}>} \text{ and } R_0^u = T \frac{\sum_{jkm} m(k+m-1) p_{jkm}}{<k_{un}>}$$ respectively, where $T$ is the average disease transmissibility and the second term is the average out-degree plus the average undirected-degree of a vertex that has become infected along a randomly selected edge. When we do not know anything about the transmission event that led to the initial infection, then our best estimate is

$$R_0 = T \frac{\sum_{jkm} (j(k+m) + m(k+m-1)) p_{jkm}}{<k_{in}> + <k_{un}>}.$$

Since $R_0$ is a product of both transmissibility (T) and the connectivity of the population, for a given value of T, different populations (networks) may have different values of $R_0$. If we are given range of $R_0$ values for a certain population, $p < R_0 < q$ for example, we can derive the lower and upper bounds for transmissibility that correspond to that range of $R_0$ as follows. Assuming we have no further information about the vaccination status of the population, we take the value of T that yields $R_0 = p$ for the population (network) with no vaccination (a worse-case scenario) and we take the value of T that yields $R_0 = q$ for the population (network) with maximum vaccination coverage (a best-case scenario.) For interpandemic flu, $R_0$ has been estimated to be $1 < R_0 < 2.4$. For an unvaccinated population, $R_0 = 1$ corresponds to $T = 0.06$ and for a population with maximum vaccination coverage (13%), $R_0 = 2.4$ corresponds to $T = 0.26$. Thus we estimate the transmissibility of interpandemic influenza to be $0.06 < T < 0.26$.

*The average size of small outbreaks and the epidemic threshold*: By nesting PGFs for the number of new infections emanating from an infected vertex one can construct a PGF for the size of a small outbreak, and hence derive the average size of a small outbreak:

$$\langle s \rangle = 1 + \frac{Tf_1^j\left[1 - T\left(f_m^{m(m-1)} - f_j^{jm}\right)\right] + Tf_1^m\left[1 - T\left(f_j^{jk} - f_m^{km}\right)\right]}{\left(1 - Tf_j^{jk}\right)\left(1 - Tf_m^{m(m-1)}\right) - T^2 f_j^{jm} f_m^{km}}$$

where $f_b^a = \dfrac{\sum_{jkm} a p_{jkm}}{\sum_{jkm} b p_{jkm}}$. When $T$ is small, the average size of a small outbreak is finite, but $\langle s \rangle$ grows with increasing transmissibility, until it diverges when the denominator of the expression above reaches its first zero. This point marks the phase transition at which the typical outbreak ceases to be confined to a finite number of cases and expands to a large-scale epidemic covering most of the network. This transition happens when $T$ is equal to the critical transmissibility $T_c$, given by

$$T_c = \frac{\left(f_j^{jk} + f_m^{m(m-1)}\right) \pm \sqrt{\left(f_j^{jk} + f_m^{m(m-1)}\right)^2 - 4\left(f_j^{jk} f_m^{m(m-1)} - f_j^{jm} f_m^{km}\right)}}{2\left(f_j^{jk} f_m^{m(m-1)} - f_j^{jm} f_m^{km}\right)}.$$

*The expected size of a full-blown epidemic $S_e$*: We can compute the size of the epidemic (the proportion of the population infected), $S_e$, for the case when $T$ is larger than $T_c$. We first calculate the likelihood that infection of a randomly chosen individual will spark only a limited outbreak instead of a full-blown epidemic, and then take one minus that probability:

$$S_e = 1 - \sum_{jkm} p_{jkm}(1+(a-1)T)^j(1-(b-1)T)^m,$$

where *a* and *b* are the solutions to the self-consistent equations

$$a = \frac{\sum_{jkm} kp_{jkm}(1+(a-1)T)^j(1+(b-1)T)^m}{<k_{out}>} \text{ and } b = \frac{\sum_{jkm} mp_{jkm}(1+(a-1)T)^j(1+(b-1)T)^{m-1}}{<k_{un}>}.$$ We use numerical root finding methods (such as Newton's method) to solve for *a* and *b*.

*The probability of a full-blown epidemic $P_e$*: The expression for $P_e$ comes from first calculating the likelihood that a single infection will lead to only a small outbreak instead of a full-blown epidemic, and then and then taking one minus the probability:

$$P_e = 1 - \sum_{jkm} p_{jkm}(1+(\alpha-1)T)^k(1-(\beta-1)T)^m,$$

where $\alpha$ and $\beta$ are the solutions to the self-consistent equations

$$\alpha = \frac{\sum_{jkm} jp_{jkm}(1+(\alpha-1)T)^k(1+(\beta-1)T)^m}{<k_{in}>} \text{ and } \beta = \frac{\sum_{jkm} mp_{jkm}(1+(\alpha-1)T)^k(1+(\beta-1)T)^{m-1}}{<k_{un}>}.$$ We use numerical root finding methods (such as Newton's method) to solve for $\alpha$ and $\beta$.

*The probability that an individual will be infected during an epidemic $v_{jm}$*: The likelihood that an individual of in-degree *j* and undirected-degree *m* will be infected during an epidemic is equal to one minus the probability that none of his or her *j + m* contacts will transmit the disease to him or her. The average probability that an individual at the (origin) end of a randomly selected directed edge is spared by an epidemic is *a*. For an individual at the end of a randomly selected undirected edge, this probability is *b*. Thus, the probability that one is not infected by a neighbor is the probability that the neighbor is infected but does not transmit disease *(1-a)(1-T)* for directed edges and *(1-b)(1-T)* for undirected edges, plus the probability that the neighbor is not infected, *a* for directed edges or *b* for undirected edges. These probabilities sum to *(1-T+Ta)* and *(1-T+Tb)* for directed and undirected edges, respectively. Thus, a randomly chosen vertex of in-degree *j* and undirected-degree *m* will become infected with probability

$$v_{jm} = 1 - (1 - T + Ta)^j (1 - T + Tb)^m.$$

*Demographic-Specific Attack Rates*: We calculate demographic-specific epidemiological risks by combining demographic information (age, occupation, etc.) for each member of the population with the $v_{jm}$, defined above. We first divide the population into 14 demographic groups:

| Demographic Group (g) | Demographic Group Description |
|---|---|
| 1 | Infants (age < 3) |
| 2 | Toddlers (3 ≤ age < 5) |
| 3 | Children (5 ≤ age < 18) |
| 4 | Adults (18 ≤ age < 50) |
| 5 | Elderly ( age > 50) |
| 6 | Nursing home residents |
| 7 | Infants in daycare |
| 8 | Toddlers in preschool |
| 9 | Health care workers |
| 10 | Nursing home workers |
| 11 | Day care workers |
| 12 | Preschool workers |
| 13 | Teachers (and school staff) |
| 14 | Unemployed |

For each demographic group (*g*), we find the expected number of infections ($N_g$) at a particular transmission probability *T* by summing the probabilities of infection ($v_{jm}$) across all individuals in that group. We denote the in-degree and undirected degree of an individual (*i*) by *j(i)* and *m(i)*, respectively:

$$N_g = \sum_{i \in g} v_{j(i)m(i)} \qquad \forall g \in [1,14].$$

*Age-Specific Mortality:* The predicted number of deaths in the population caused by an epidemic (*M*) is the product of the predicted number of infections in each of the age groups (demographic groups 1 through 5 in the Table above) and the age-specific mortality rate ($R_g$) specified in Table 3:

$$M = \sum_{g=1}^{5} N_g * R_g$$

The predicted total mortality rate for the population is *M* normalized by the population size.

*Multiple Introductions*: We can also analytically predict the probability of an epidemic given independent multiple introductions of disease into a population. For a given number of introductions, *n*, the probability of an epidemic is given by:

$$\pi_n = 1 - (1 - P_e)^n,$$

where, $P_e$, is the probability of an epidemic assuming a single introduction. We note that the calculation of $\pi_n$ above assumes that all *n* introductions occur independently at the outset of an outbreak. This assumption yields an upper bound estimate for the probability of an epidemic with multiple introductions.

*Epidemic Simulation*

We verify the analytic predictions using simulations of a Susceptible-Infectious-Recovered (SIR) model. The simulations are initialized with an entirely susceptible population, except for a single infected case (*patient zero*). An infected vertex passes the disease on to each of its neighbors (those with whom that individual has disease-causing contacts) with probability *T* (the average transmission probability). This process continues until the population no longer includes any susceptible individuals that are in contact with any infected individuals. Once an individual has had the chance to infect its neighbors, it is immediately moved into the recovered class.

SUPPLEMENTAL ANALYSIS

*Network Properties of Demographic Groups*

Here we describe basic properties of the simulated urban networks that we have analyzed. The epidemiological calculations consider the degree distribution of the network (as described in the previous section.) Recall that most of the edges in our network are undirected and many individuals have the same out-degree as in-degree, with the exception of health care workers and individuals who are at high risk for complications due to flu. In Figure S1, we show the in-degree distributions for the total population and select demographic groups before and after vaccination by the morbidity and mortality-based strategies. Children have a much higher mean in-degree (24.1) than adults and elders (10.7 and 10.6, respectively). Figure S1c illustrates that the contact patterns for adults are relatively unaffected by both the morbidity- and mortality-based strategies. The morbidity-based strategy primarily alters the degree distribution of children (Figure S1b) and the mortality-based strategy primarily alters the degree distribution of elders (Figure S1d). The mortality-based strategy does not affect the degree distribution of the total population a great deal (Figure S1a) as it effectively only targets small groups, either due to low vaccine efficacy levels (elders) or few individuals in the demographic group (infants.)

*Sensitivity to Population Structure*

The urban networks are stochastically generated, yielding Poisson distributions of contact numbers within each setting (schools, hospitals, workplaces, etc.). To achieve this, we specify setting-specific probabilities that determine whether or not any given pair of individuals in the same location will have an edge drawn between them. We examined the sensitivity of our results to the specific probabilities used in generating the network. First, we generate 100 networks each with 5000 households. (The smaller population size allowed for more extensive sensitivity analyses. In prior studies, we found that epidemiological predictions for small urban networks apply to large urban networks [S5], and thus we expect that these sensitivity results will also apply to large networks.) To generate variation in these networks, we draw contact probabilities between individuals from a Gaussian distribution, and allow them to deviate by up to 100% from the original contact parameters (which range from 0.003 to 0.3 depending on the location/nature of the interaction). The stochastic formation of edges according to these probabilities yields 100 unique networks, each with its own degree distribution. We then vaccinate each population according to the morbidity-based and mortality-based strategies, as described in the methods section. The dashed lines in Figure S2 indicate the standard deviations for each epidemiological prediction (morbidity and mortality) across the 100 networks. The small variation in the predictions indicates that our results are robust to stochastic variation in network structure. Even with a 100% deviation in contact structure, the morbidity and mortality-based strategies are superior for lower and higher values of $T$, respectively. The value of $T$ at which preferred strategy switches falls in the range [0.10, 0.14]. In the main text, we report a transition point of $T$=0.11. These results suggest that even a 100% uncertainty in contact rates produces an uncertainty of [-0.01, 0.03] around the transition point.

*Sensitivity to Mortality Rates*

To evaluate the sensitivity of our predictions to variations in virulence among different strains of influenza, we evaluated vaccination strategies for two markedly different estimated mortality distributions. Here we extend this analysis to several other estimated influenza mortality distributions. We compare the total mortality caused by an influenza epidemic after the population has been vaccinated with a morbidity-based or mortality-based strategy for five different age-specific mortality distributions. The first two mortality distributions are the focus of the main text. The third is a different estimate for 1918 flu mortality rates that includes high mortality rates for adults and the elderly; and the remaining two are U-shaped mortality distributions reported for the epidemics of 1892 and 1936 [S8]. The mortality-based strategy is designed to target the demographic groups with the highest mortality rates, and thus varies from one mortality distribution to the next. The morbidity-based strategy is identical across all five mortality distributions, targeting school children and staff as specified in the main text. We are particularly interested the cross-points (in transmissibility values) where the mortality-based strategy becomes superior to the morbidity-based strategy. These lie between $T = 0.15$ and $T = 0.20$ for the three additional mortality distributions shown in Figure S3, very close to that predicted for the 1918 mortality distribution considered in the main text, further suggesting that the results are fairly insensitive to uncertainties in estimates of influenza mortality rates.

*Sensitivity to Vaccine Coverage Level*

We test the sensitivity of our results to a change in the vaccine coverage level. The vaccine priorities in the main text are implemented at a 13% coverage level. Here, we implement the morbidity-based strategy (school children and staff) and mortality-based strategy (based on the second mortality distribution of Figure S3) at a 20% coverage level. (During the influenza vaccine shortage of 2004, enough vaccine was available to cover 20% of the population.) Figure S4 shows that the increase in vaccination coverage produces a smaller mortality rate for both strategies but the comparison is qualitatively similar: the mortality-based strategy outperforms the morbidity-based strategy for higher values of transmissibility.

*Vaccine Effectiveness and Efficacy*

Vaccine efficacy is defined as $E = 1 -$ (attack rate among vaccinated population/attack rate among unvaccinated population). To model vaccination of a proportion $C$ of a demographic group at an efficacy of $E$, we remove a fraction $C*E$ of individuals in the group from the network entirely. This reduces the attack rate in the vaccinated group by exactly a fraction $E$, which yields an efficacy of $E$. Although this method technically assumes that the vaccine has 100% effectiveness for the fraction $E$ of the vaccinated group and no effectiveness on the remaining fraction $1-E$ of vaccinated individuals, it provides a reasonable model for the more realistic scenario in which most vaccinate individuals enjoy some level of protection. To test that the 100% effectiveness model is a reasonable approximation, we have compared its predictions to simulations in which all vaccinated individuals have partially reduced susceptibility. In particular, to vaccinate a fraction $C$ of a group with a vaccine of efficacy $E > 0$, we select $C$ individuals at random from the group and reduce the transmission probability along all edges leading to each of those individuals by a factor $(1-E^{1/m})/T(1-b)$, where $m$ is the undirected degree of the node, $T$ is the average transmissibility, and $b$ is the average probability that an individual at the end of a randomly selected undirected edge is spared by an epidemic. This factor is based on the fact that the probability of infection for each vaccinated individual will be *1-E*, and this yields a reduction in the attack rate in the group of $E$. We then simulate the spread of disease through the network. We find that mathematical predictions assuming 100% effectiveness closely match the results of these simulations, as illustrated in Figure S5.

*Sensitivity to Variation in Infectivity and Susceptibility*

There is certainly heterogeneity in influenza infectivity and susceptibility among individuals. Some of the heterogeneity is caused by variation in contact patterns [S9]. Individuals with more contacts will have greater opportunities to catch and spread disease. Our models explicitly capture this source of variation. The remaining heterogeneity in transmission rates is caused by intrinsic physiological and behavioral differences among individuals. Our analytical calculations allow for such heterogeneity so long as it is distributed somewhat randomly with respect to the structure of the population. That is, there should not be significant correlations between individual contact patterns and individual likelihoods of infection and/or transmission. There is evidence, however, that such correlations may exist. Cauchemez et al. statistically argue that children have a higher infectiousness and a higher susceptibility than adults *per contact* [S10].

We have modified our contact network to explicitly model this diversity in transmissibility. Cauchemez et al. suggest that, within a household, susceptible children (under 15 years of age) have a 15% higher susceptibility per day of contact with an infectious household member compared to susceptible adults. They estimate that infectious children have 84% higher infectivity per day of contact with susceptible household members compared to infectious adults (0.26 and 0.48 for adults and children, respectively). Finally, they find that the expected infectious periods of flu for children and adults are 3.6 and 3.9 days, respectively [S10].

To test the sensitivity of our model to such heterogeneity, we make the extreme assumption that the different transmission probabilities will also hold for contacts outside the home. Accordingly, contacts between children were given the highest probability of transmission ($T_{CC}$), followed by those from children to adults ($T_{CA}$), from adults to children ($T_{AC}$), and finally between adults ($T_{AA}$). We assigned probabilities of transmission to the edges in the network based on the ratios between these values calculated from the data reported in [S10]. In particular, each transmissibility is the average probability of transmission between an infectious and susceptible individual during the infectious period, or $1-(1-p)^\tau$ where $\tau$ is the length of the infectious period and $p$ is the per day probability of infection. Borrowing notation from [S10], for an edge pointing from an infected individual of type $I$ (child or adult) to a susceptible individual of type $J$ (child or adult) this is equal to $1-(1-\beta_I \varepsilon_J)^{\tau_I}$ where $\beta_I$ is the per day probability of transmission from $I$ to one of his or her contacts, $\tau_I$ is the duration of $I$'s infectious period, and $\varepsilon_J$ is the susceptibility factor of individual $J$. The table below gives the ratios that we used to determine transmissibilities in the model.

| Contact Type ($I \rightarrow J$) | Transmissibility calculated from estimates reported [S10] $1-(1-\beta_I \varepsilon_J)^{\tau_I}$ | Ratio of $T_{IJ}$ to $T_{AA}$ |
|---|---|---|
| Adult $\rightarrow$ Adult | $1-(1-0.26)^{3.9}$ | 1 |
| Adult $\rightarrow$ Child | $1-(1-0.26 \cdot 1.15)^{3.9}$ | 1.09 |
| Child $\rightarrow$ Adult | $1-(1-0.48)^{3.6}$ | 1.31 |
| Child $\rightarrow$ Child | $1-(1-0.48 \cdot 1.15)^{3.6}$ | 1.37 |

For a range of possible values for $T_{AA}$, we assigned transmissibilities according to these ratios and ran 250 SIR simulations on the network for each of three scenarios: no vaccination, mortality-based vaccination, and morbidity-based vaccination. We also calculated the average transmissibility $T$ across all contacts in the network, with which we made analytical predictions. In Figure S6, we compare analytical predictions that consider only the average transmissibility $T$ (lines) to the outcome of these simulations (circles). Our current analytic methods give qualitatively similar results to those of the simulations. Although there is a bit of a discrepancy between the predicted and simulated results for the morbidity-based strategy and consequently for the transition point between the two strategies, the important results still hold. That is, the mortality-based strategy remains advisable over a large range of highly contagious strains (even larger in the simulations than the analytics), and the cross-point between the two strategies remains within the range of estimates of $R_0$ for interpandemic flu.

Although the results in Figure S6 may seem counter-intuitive, they point to some of the important features of our model. Generally, the morbidity-based strategy reduces mortality via herd immunity while the mortality-based strategy reduces mortality by directly protecting those with highest mortality rates. By assuming that edges to and from children have higher transmissibilities, it becomes more difficult to achieve herd immunity via the morbidity-based strategy. Despite the fact that the morbidity-based strategy targets a core group (50% of all children), it leaves a substantial population of children with high degree and high infectiousness which continues to act as a core group sufficient enough to reach the high-risk individuals. The mortality-based strategy is relatively unaffected by the heterogeneity because it continues to protect the same fraction of high-risk individuals (in neither case does it achieve much herd immunity).

This sensitivity analysis was based on a fairly extreme form of heterogeneity in transmission probabilities. In reality, variation in transmission probabilities outside households may be less demographically-structured, in which case, the assumption that variation in transmission probabilities is fairly random with respect to network structure may be valid. Similar analytic methods that explicitly capture demographic-specific patterns of transmission rates give more exact predictions for this extreme scenario, but are beyond the scope of this paper.

**References**

S1. Household Size, census metropolitan areas. Statistics Canada (2001)

S2. 2001 Census Profile of British Columbia's Regions: Greater Vancouver Regional District, *BC Stats* (2003)

S3. Vancouver School Board, December 2002 Ready Reference (2002).

S4. The British Columbia Health Atlas, Centre for Health Services and Policy Research (2002)

S5. Pourbohloul B, Meyers LA, Skowronski DM, Krajden M, Patrick DM, et al. (2005) Modeling Control Strategies of Respiratory Pathogens. Emerg Infect Dis. 11: 1249-1256

S6. Meyers LA, Newman MEJ, Pourbohloul B (2006) Predicting epidemics on directed contact networks. JTB. In press.

S7. Newman MEJ (2002) Spread of epidemic disease on network. Phys. Rev. E 66, 016128

S8. Dauer CC, Serfling RE (1961) Mortality from Influenza, 1957-1958 and 1959-1960. Am Rev Respir Dis. 83 (2 Suppl): 15-26.

S9. Addy CL, Longini IM, Harber M (1991) A Generalized Stochastic Model for the Analysis of Infectious Disease Final Size Data. Biometrics. 47: 961-974.

S10. Cauchemez S, Carrat F, Viboud C, Valleron AJ, Boelle PY (2004) A Bayesian MCMC approach to study transmission of influenza: application to household longitudinal data. Stat Med. 23: 3469-3487.

S11. Thompson WW, Shay DK, Weintraub E, Brammer L, Cox N et al. (2003) Mortality associated with influenza and respiratory syncytial virus in the United States. JAMA. 289: 179–86.

S12. Simonsen L, Clarke MJ, Schonberger LB, Arden NH, Cox NJ et al. (1998) Pandemic versus epidemic influenza mortality: a pattern of changing age distribution. J Infect Dis. 178: 53–60.

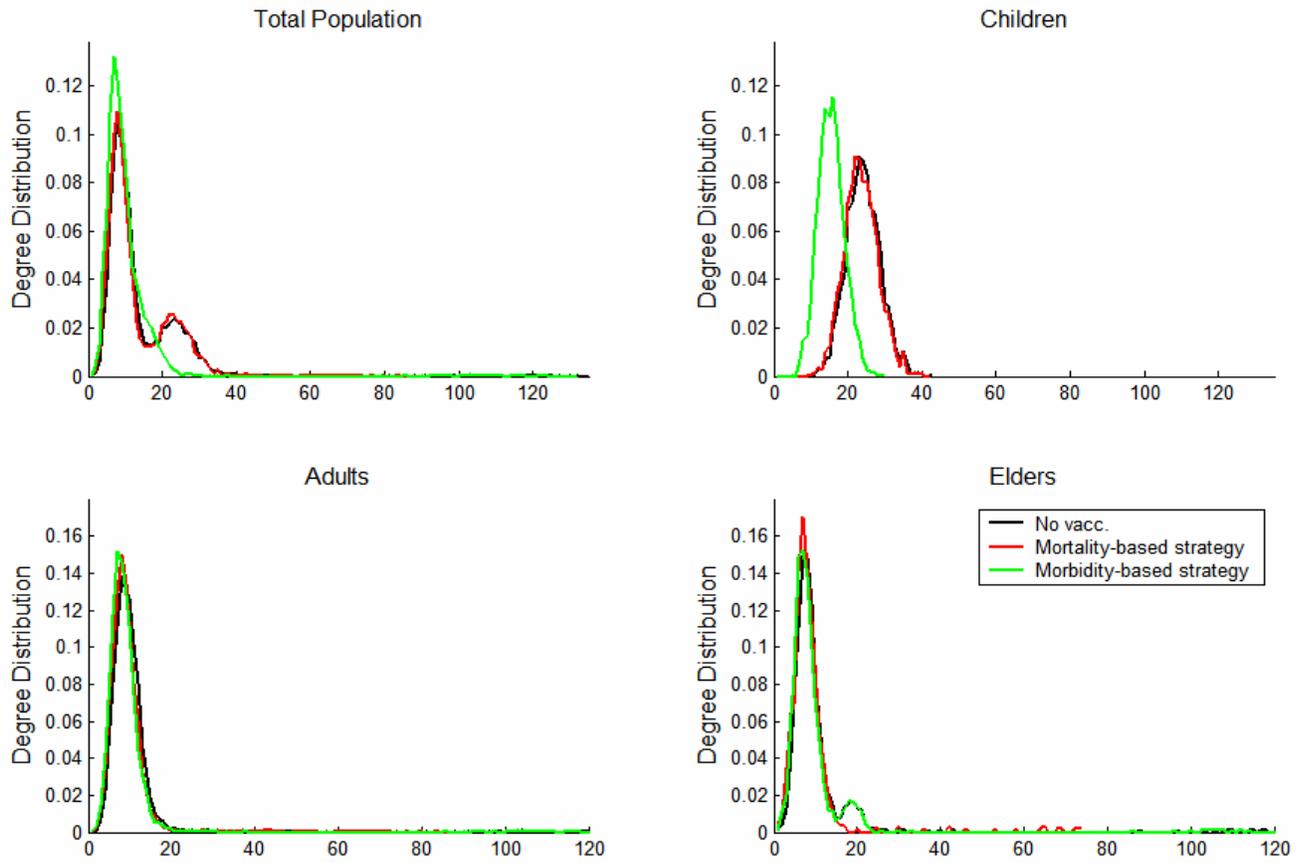

Figure S1: (Normalized) Degree distributions for various demographic groups before and after vaccination (with the interpandemic mortality-based strategy and the morbidity-based strategy.) Vaccinated individuals are not included in the distributions shown for the two strategies.

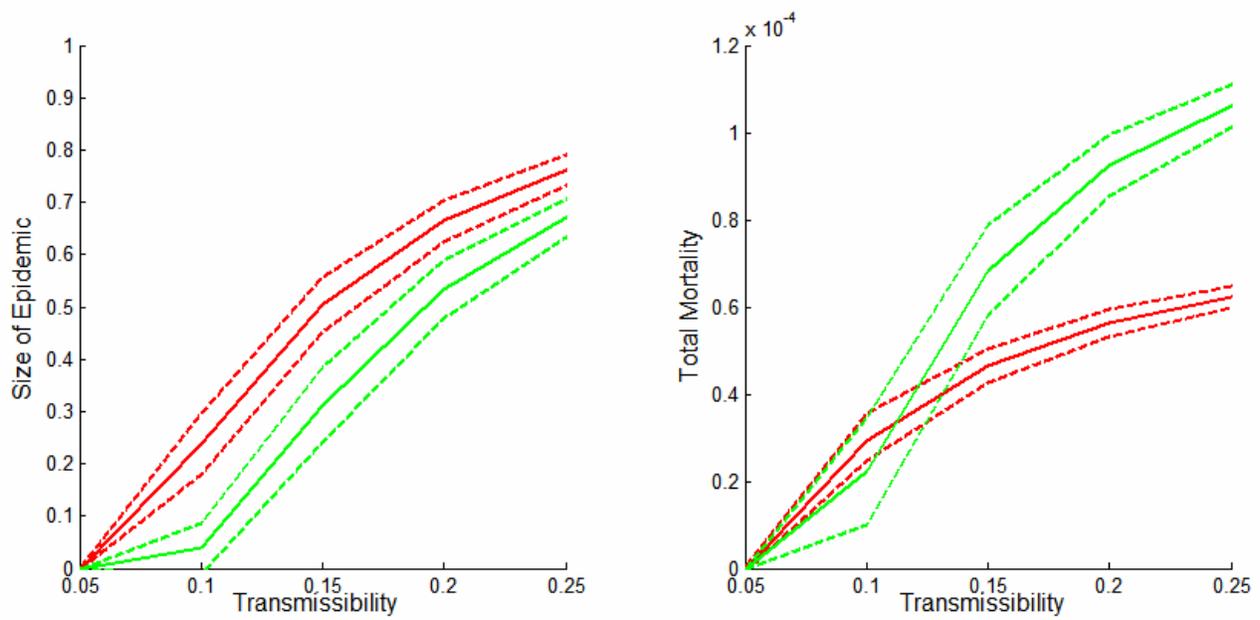

Figure S2: Variation in the size of epidemic and total mortality predicted for mortality-based (red) and morbidity-based (green) strategies across 100 networks with 100% variation in contact parameters. Dashed lines indicate standard deviations.

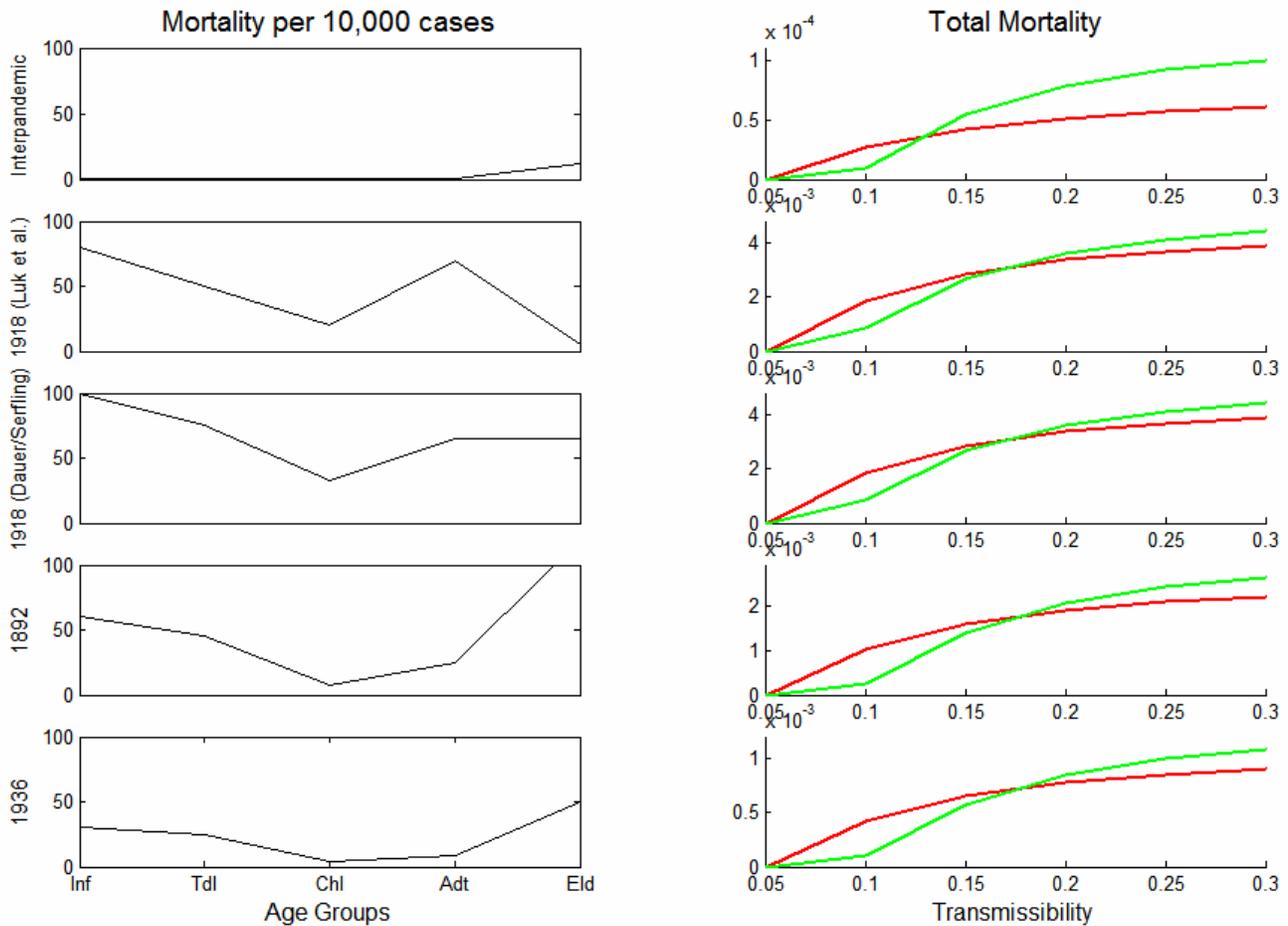

Figure S3: Epidemiological predictions for five different estimated influenza mortality distributions. Left: Mortality rates estimated for various influenza epidemics and pandemics [S8, S11, S12]. The top two distributions are considered in the main text. Right: Total mortality predicted for the mortality-based strategy (red) and morbidity-based strategy (green) assuming the corresponding distributions of mortality rates.

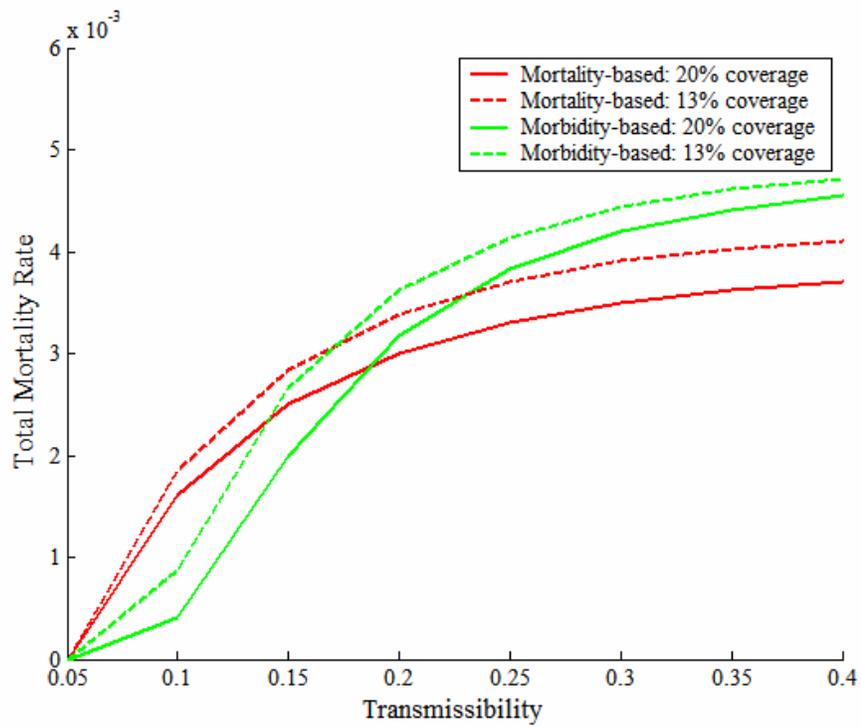

Figure S4: The total mortality at a 20% vaccination coverage level. The total mortality is lower for both strategies as compared to vaccination at a 13% coverage level (dashed lines). However, there still exists a point after which the mortality-based strategy is superior to the morbidity-based.

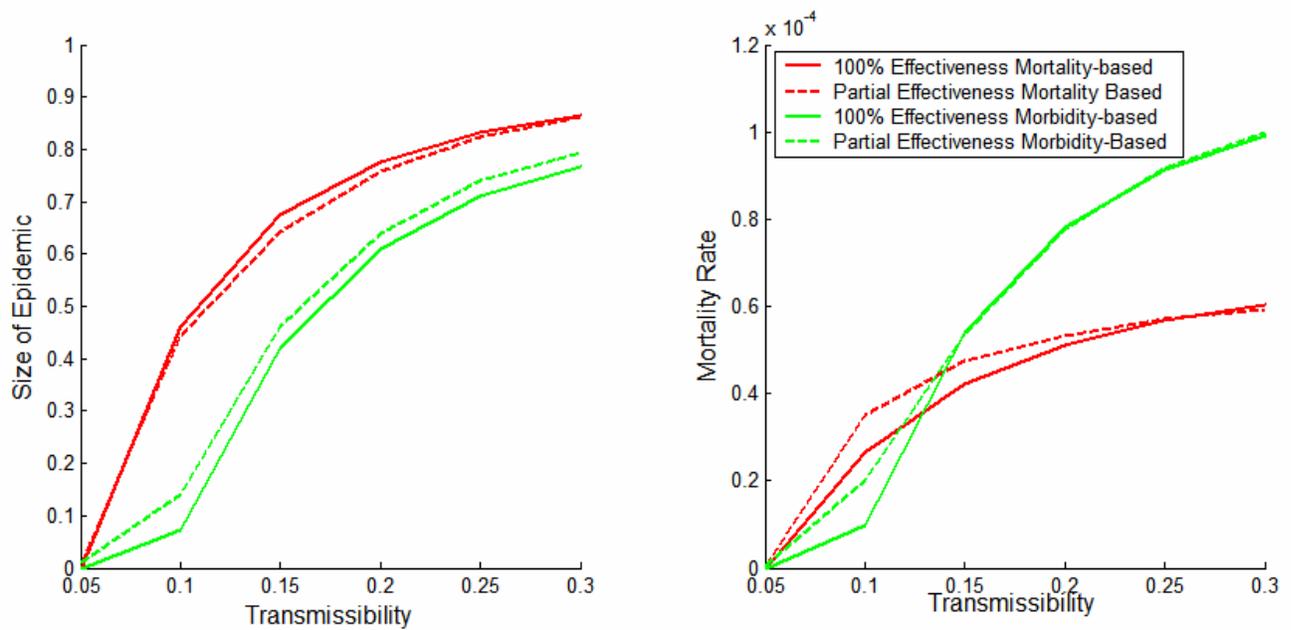

Figure S5: Results from simulation demonstrate that the two methods of modeling vaccine efficacy give similar results. Results are shown for the mortality-based and morbidity-based strategies. 'Partial effectiveness' refers to the vaccination of a proportion $C$ of the population with effectiveness $E$. '100% effectiveness' refers to the vaccination of a proportion $C*E$ of the population with effectiveness 1.

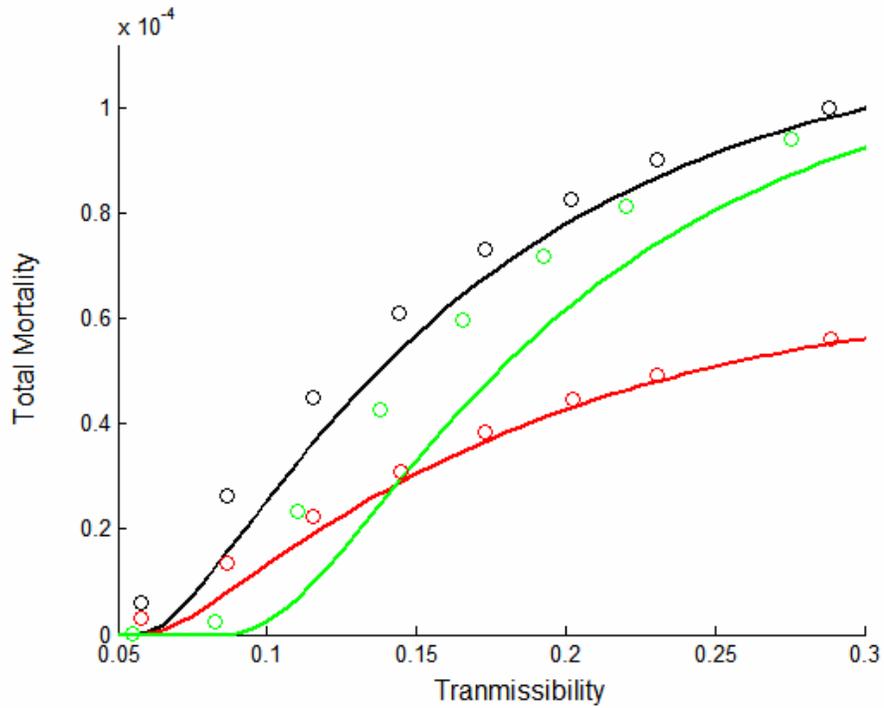

Figure S6: Results for total mortality rate with variation in infectivity and susceptibility. The x-axis corresponds to the average transmissibility across all edges in the network. Circles are simulation results with individual variation in infectivity and susceptibility; lines are analytical results for the resulting average transmissibility on the same networks. The analytical calculations do not explicitly consider variation in infectivity and susceptibility on each edge.